\documentclass[]{aa}       
\usepackage{graphicx}
\usepackage{amssymb}
\usepackage{longtable}
\usepackage{supertabular}
\usepackage{natbib} 
\usepackage{ulem}

\begin{document}

\title{Clumpy stellar winds and high-energy emission in high-mass binaries hosting a young pulsar}

\author{
V. Bosch-Ramon \inst{1}
}

\authorrunning{Bosch-Ramon, V.}

\titlerunning{Clumpy stellar winds in young pulsar high-mass binaries}

\institute{
Departament d'Astronomia i Meteorologia, Institut de Ci\`encies del Cosmos (ICC), Universitat de Barcelona (IEEC-UB), Mart\'i i
Franqu\`es 1, 08028 Barcelona, Spain, e-mail: vbosch@am.ub.es
}

\offprints{V. Bosch-Ramon, \email{vbosch@am.ub.es}}

\date{Received <date> / Accepted <date>}

\abstract
{High-mass binaries hosting young pulsars can be powerful gamma-ray emitters. The stellar wind of the massive star in the system, which interacts with the pulsar wind, is expected to be clumpy. Since the high-energy emission comes from the  interaction of the two winds, the presence of clumps can affect the spectrum and variability of this radiation.}     
{We look for the main effects of the presence of clumps in the stellar wind on the two-wind interaction region and on the non-thermal radiation that originates there.}    
{A simple analytical model for the two-wind interaction dynamics was developed. The model accounts for the lifetime of clumps under the pulsar-wind impact. This time plays a very important role with regard to the evolution of the clump, the magnetic field in the clump-pulsar wind interaction region, and the non-radiative and radiative cooling of the non-thermal particles. We also computed the high-energy emission produced at the interaction of long-living clumps with the pulsar wind.}    
{For reasonable parameters, the clumps will induce small variability on the X-ray and gamma-ray radiation. Sporadically, large clumps can reach closer to the pulsar increasing the magnetic field, triggering synchrotron X-ray flares and weakening other emission components like inverse Compton. The reduction of the emitter size induced by clumps also makes non-radiative losses faster. Stellar wind clumps can also enhance instability development and matter entrainment in the shocked pulsar wind when it leaves the binary.
Growth limitations of the clumps from the wind acceleration region may imply that a different origin for the largest clumps is required. The large-scale wind structures behind the observed discrete absorption components in the ultraviolet may be the source of these large clumps.}  
{The presence of structure in the stellar wind can produce substantial energy-dependent variability and thus should not be neglected when studying the broadband emission from high-mass binaries hosting young pulsars.} 

\maketitle

\section{Introduction}\label{intro}

Several binary systems in the Galaxy are powerful gamma-ray emitters \citep[see][and references therein]{2013APh....43..301P,2013arXiv1307.7083D}. Among these binaries, PSR~B1259$-$63(/LS~2883) is a system formed by a late O star \citep{2011ApJ...732L..11N} and a powerful, relatively young, non-accreting pulsar \citep{1992ApJ...387L..37J}. The remaining gamma-ray emitting binaries, with the exception of the X-ray binaries Cygnus~X-1 (to be confirmed) and Cygnus~X-3, are objects of unclear nature that may also belong to the same class as PSR~B1259$-$63. 

Young-pulsar high-mass binaries are not expected to be very numerous in the Galaxy because of the short duration of the pulsar non-accreting phase compared with the whole life-span of massive X-ray binaries with neutron stars. In addition to PSR~B1259$-$63, only a few less powerful systems lacking non-thermal radiation have been found so far (see Table 4 in  \citealt{2012ApJ...756..188P}). Assuming a lifetime for the non-accreting pulsar of a few times $10^5$~yr (the age of PSR~B1259$-$63), and the birth rates for high-mass binaries hosting a neutron star from \cite{1999MNRAS.309...26P}, one might expect $\sim 100$ of those systems in the Galaxy. Despite their relative scarcity, however, given the complex high-energy phenomena that arise from the conjunction of the orbital motion, the presence of a massive star, and the interaction of the stellar and the pulsar winds, these objects are important for the study of pulsar and stellar wind physics, particle acceleration, supersonic and relativistic fluid dynamics, emission mechanisms, and radiation reprocessing. 

Important physical information can be extracted from young-pulsar high-mass binaries through observations in different energy bands and spectral, variability and morphological studies, although this requires a proper understanding of the main factors that come into play. In this regard, since the interaction region between the stellar and the pulsar winds is a natural site for non-thermal radiation \citep[e.g.][etc.]{1981MNRAS.194P...1M,1997ApJ...477..439T,1999APh....10...31K,2006A&A...456..801D,2007MNRAS.380..320K,2007Ap&SS.309..253N}, the stellar wind properties are to be properly characterized for a sensible modelling of the dynamics of the interacting flows and their emission. 

Concerning the non-thermal emission from the two-wind interaction region, it is expected that the gamma-ray production will be mostly affected by dynamical processes taking place close to or within the binary (e.g. \citealt{2007MNRAS.380..320K}; see also \citealt{2013A&A...551A..17Z}), whereas observations show \citep[see][]{2011ApJ...732L..10M} that radio emission is produced much farther from the system \citep[see also][]{2006A&A...456..801D,2010tsra.confE.179B}. Therefore, the structure of the stellar wind and its impact on the non-thermal processes should be better probed at high energies at which the radiation from the two-wind interaction region can present the fastest and sharpest variations.

The structure of massive star winds is characterized by the presence of density and velocity irregularities or clumps \citep[e.g.][]{2002A&A...381.1015R}. In the past, numerical and analytic studies of the influence of stellar wind clumping on thermal and non-thermal radiation have been performed for massive star binaries \citep[e.g.][]{2007ApJ...660L.141P}, high-mass microquasars 
\citep[e.g.][]{2009ApJ...696..690O,2009A&A...503..673A,2012A&A...539A..57P}, and LS~I~+61~303 \citep[e.g.][]{2010MNRAS.403.1873Z}, a high-mass binary candidate that may host a non-accreting pulsar. So far, however, a specific investigation of the characteristic timescales and broadband flux variations introduced by the presence of stellar wind clumps in young pulsar high-mass binaries has not been done. In what follows, we present a simplified treatment of this problem to evaluate the observational impact of the arrival of clumps to the two-wind interaction region at X- and gamma rays. This study can provide a framework for more detailed calculations.

\section{The clumpy stellar wind}\label{clw}

The winds of massive stars are produced through radiation pressure exerted via the interaction of ultraviolet (UV) photons with electrons bound to atoms in the stellar atmosphere \citep{1970ApJ...159..879L}. Since the acceleration of the wind is a highly unstable process, it leads to strong irregularities in density and velocity \citep{2002A&A...381.1015R}. These irregularities, often referred to as clumps, have an impact on the observables of massive star winds, in particular affecting the inferred values of the mass-loss rates, which are crucial for the understanding of stellar evolution 
\citep[e.g.][]{2008cihw.conf....9H}. The characteristics of clumps are still unclear, but it is thought that they carry most of the wind mass, but fill only a small fraction ($f<1$) of the volume in the stellar surroundings, yielding a contrast  of $f^{-1}$ in density with respect to its average value. The clump sizes are expected to be much smaller than the star  
\citep{2006ApJ...648..565O}, although larger clumps may also be present in the wind like the clumps used, for instance, to explain the X-ray flares in supergiant fast X-ray transients  \citep[e.g.][]{2007A&A...476..335W}, or the large scale structures behind the observed discrete absorption components in the UV in early-type stars (DACs) \citep[e.g.][]{1996ApJ...462..469C,2008ApJ...678..408L}. The clumps should present a distribution of masses and sizes if they are the product of non-linear processes at the wind base \citep{2008cihw.conf...17M}. 
In this investigation, clumps with different masses and sizes are empirically considered. The values discussed are representative of the majority of the clumps in the wind, i.e. the { \it average} clump case, or exceptional, i.e. the arrival of a rare and large matter condensation.

\section{The two-wind interaction region}

\subsection{Physical framework}\label{phfr}

\begin{figure}
\includegraphics[width=0.5\textwidth,angle=0]{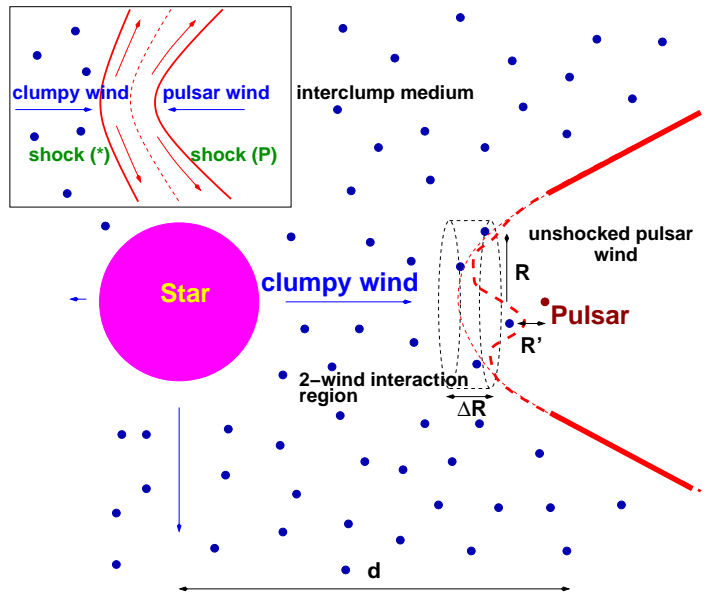}
\caption{Schematic of the scenario considered in this work: a clumpy stellar wind interacts with the pulsar wind forming a two-wind interaction region that varies in space and time because of the arrival of dense and/or large clumps. In the top-left corner of the figure, a zoom in of the colliding two-wind region is shown. }
\label{sketch}
\end{figure}

Stellar and pulsar winds collide forming two shocks and a region in between filled with shocked wind material. These two shocks contain a contact discontinuity that is located where the ram pressures of the two winds balance each other. The two-wind interaction structure becomes bow-shaped towards the wind with the lower momentum flux.  In the top-left corner of Fig.~\ref{sketch}, a representation of the colliding two-wind region is shown.

The shocked stellar and pulsar winds flow in contact but at very different speeds, suffering re-acceleration \citep{2008MNRAS.387...63B}, the Kelvin-Helmholtz instability, and mixing, as they leave the binary system \citep{2013arXiv1309.7629L}. All these phenomena, and the orbital motion, play a major role in shaping the interaction structure at larger scales \citep{2011PASJ...63..893O,2012A&A...546A..60L,2012A&A...544A..59B}, and possibly the highest energy emission as well \citep[see][]{2013A&A...551A..17Z}. In addition, a clumpy stellar wind can have consequences on the shocked flow dynamics and the radiation formation processes that might be significant and should not be overlooked. 

Here, we restrict our study to the region confined within the binary system. In Fig.~\ref{sketch}, a schematic of the considered scenario is presented.  For simplicity, in what follows we consider the two-wind interaction region to be a circular slab with radius $\sim R$ and thickness $\Delta R$. The solid angle of the clump-pulsar wind interaction region, as seen from the pulsar, is $\Delta \Omega\approx \pi(R_{\rm c}/R)^2$. The quantities $R$, $R_{\rm c}$, and $\Delta R$ are the distance between the two-wind interaction region and the pulsar, the clump radius, and the thickness of the shocked two-wind zone, respectively. We neglect here the bow shape of the two-wind and clump-pulsar wind shocks, as well as the shocked material flowing beyond the interaction slab. Under all these approximations, the escape timescale of the material in the shocked wind zones, similar to the local adiabatic cooling time, can be taken as $R/u$ and $R_{\rm c}/u$ in the two-wind and the clump-pulsar wind shocks, respectively. The flow velocity $u$ will be taken as the sound speed, similar to the stellar wind (and clump) velocity $V_{\rm w}$, in the shocked stellar wind, and the relativistic sound speed $c/\sqrt{3}$ in the shocked pulsar wind. 
In our study, the wind and its clumps always interact with the pulsar wind well beyond the pulsar magnetosphere.

The properties of the two-wind and the clump-pulsar wind interaction regions are assumed to be uniform, which will greatly simplify the radiation treatment in Sect.~\ref{rad}. Another simplification of the emitting regions done here is to consider them to be point-like (see Sect.~\ref{semian}). Finally, we assume that clumps arrive to the two-wind interaction region moving along the line joining the two stars. All these approximations will allow us to obtain order-of-magnitude estimates for the relevant timescales and flux variations at different energy bands, capturing the main features of the influence of the clump on the radiation. For accurate predictions, however, detailed studies of the dynamics and radiation should be combined, accounting also for the shocked flow evolution outside the binary system. 

\subsection{Relevant spatial and timescales of the problem}\label{tscl}

The observational impact of a clumpy stellar wind strongly depends on the density contrast in the two-wind interaction region, which gives the clump enhancement in momentum flux. The sound speed of the shocked stellar wind when it is smooth is $\sim V_{\rm w}$ at the region where its ram pressure is balanced by the pulsar wind, whereas the sound speed of a shocked clump at the same location gets reduced by $f^{1/2}$ because of the higher density \citep[see e.g.][and references therein]{2009A&A...503..673A}, so the time required to shock the clump when it meets the interaction region is
\begin{equation}
t_{\rm c}\approx f^{-1/2}R_{\rm c}/V_{\rm w}\,. 
\label{tc}
\end{equation}

For a clump radius such that $R_{\rm c}\ll f^{1/2}\Delta R$
the clumps will be destroyed  and deflected by the shocked wind medium well before crossing it, being quickly integrated and carried away from the binary \citep{2007ApJ...660L.141P}. Otherwise, for 
\begin{equation}
R_{\rm c}\gtrsim f^{1/2}\Delta R\,, 
\label{cond1}\end{equation}
the clumps will not be smoothed out within the shocked two-wind zone, but will penetrate into the unshocked pulsar wind. Depending on $f$ and the {\it average} clump size, the two-wind interaction region as a whole will be smooth or very inhomogeneous. In the latter case, the shape and location of the two-wind interaction region will vary in time.

Under typical pulsar-to-star wind momentum-flux ratios, say $\eta\sim 0.03-0.3$ \citep[see Sect. 1 in][]{2012A&A...544A..59B}, the total luminosity injected by the pulsar wind within the binary will be 
$\sim (2\eta\,c/V_{\rm w})$, $\sim 10-100$ times larger than that of the star. 
This suggests that the pulsar wind shock should largely dominate the non-thermal output from the two-wind interaction region. In addition, $R$ determines the magnetic field through $B\propto R^{-1}$ for a dominant toroidal $B$-component, and the cooling time $\sim \sqrt{3}\,R/c$ for escape / adiabatic (non-radiative) cooling. Thus, clump-induced variations in the pulsar wind shock could strongly affect the properties of the non-thermal emission, namely the broadband spectrum and variability. 

Considering the solid angle covered by the clump as seen from the pulsar, the luminosity of a clump-pulsar wind shock will be smaller by $\sim \min[(R_{\rm c}/R')^2,1]$ than the 
whole two-wind interaction region, where $R'=R-X$, and $X$ is the distance the clump can cover within the unshocked pulsar wind: 
\begin{equation}
X\sim t_{\rm c}V_{\rm w}\sim f^{-1/2}R_{\rm c}\,.
\end{equation}
To study the case $R_{\rm c}\sim R'$, one can write
\begin{equation}
(R_{\rm c}/R')^2\sim \left(\frac{R_{\rm c}}{R-X}\right)^{2}\sim \left(\frac{R_{\rm c}}{R-f^{-1/2}R_{\rm c}}\right)^{2}
\end{equation}
to order 1 if 
\begin{equation}
R_{\rm c}\sim R/(1+f^{-1/2})\sim f^{1/2}R\,.
\label{cond2}
\end{equation}
Equation~(\ref{cond2}) provides a stricter condition than that derived from Eq.~(\ref{cond1}) by a factor of several, given the typical thickness of the two-wind interaction region right on the line joining the two stars \citep[see e.g.][]{2008MNRAS.387...63B}. Clumps fulfilling this condition will be able to penetrate up to the location at which the pulsar wind ram pressure equals that of the clump. This yields for these long-lived clumps
\begin{equation}
R' \sim f^{1/2}R\sim R_{\rm c}\,, 
\label{cond3}
\end{equation}
which means that most of the pulsar wind will terminate in the clump. 
At that point, the clump will stop its advance and dissipate under the impact of the pulsar wind. This process is discussed in Sect.~\ref{disc}. 

\subsection{Clump size}\label{clsd}

As described in Sect.~\ref{clw}, the stellar wind clumps form through complex non-linear, radiation-hydrodynamical processes in the stellar atmosphere. Although empirical and theoretical clump characterization is still under development, clumps can be considered dense, changing structures embedded in a hotter and very dilute medium. Clumps therefore expand and evolve in the stellar wind in a complex manner. In particular, $f$ and thereby the density of the clumps, as well as $R_{\rm c}$, seem to be functions of $d$ \citep[e.g.][]{2006A&A...454..625P}, not only because clumps expand as they propagate away from the star, but also because of collisions between clumps that produce their merging or fragmentation \citep[e.g.][]{2006ApJ...648..565O}. 

In this work we are interested in those known binaries that host or may host a non-accreting pulsar \citep[see][and references therein]{2013APh....43..301P,2013arXiv1307.7083D}. These binaries are relatively compact, with periods ranging from days to years, and their orbital semi-major axis typically going from a few $\times 10^{12}$ to a few $\times 10^{13}$ cm. To investigate the impact of a clumpy wind, we will normalize $d=10\,d_{1}R_*$, where $d$ is the orbital separation distance and $R_*$ the stellar radius, fixed to $10\,R_\odot\approx 7\times 10^{11}$~cm. The parameter $f$ will be normalized at $0.1\,f_{-1}$.

The pulsar-to-star wind momentum flux ratio $\eta$ is also important here. Typically, $\eta\ll 1$ (see Sect.~\ref{tscl}), so we can write
\begin{equation}
R=\sqrt{\eta}\,d/(1+\sqrt{\eta})\approx \eta^{1/2}d\approx 3\,\eta_{-1}^{1/2}d_{1}\,R_*,
\label{Rd}
\end{equation}
where $\eta_{-1}=(\eta/0.1)$. For $\eta=0.1$ the error of the approximation is $\approx 30\%$ and roughly decreases as $\propto\eta^{1/2}$ for smaller $\eta$-values.
Equations~(\ref{cond3}) and (\ref{Rd}) provide a minimum radius for the clump to deeply penetrate into the unshocked pulsar wind:
\begin{equation}
R_{\rm c}\sim f_{-1}^{1/2}\eta_{-1}^{1/2}d_{1}R_*\,.
\label{rccwr}
\end{equation} 

The possibility of having such large structures from instabilities in the wind acceleration region deserves more careful considerations. Assuming clump-free expansion $R_{\rm c}\propto d$, Eq.~(\ref{rccwr}) implies clump sizes at the wind base of $\sim 0.1\,f_{-1}^{1/2}\eta_{-1}^{1/2}d_{1}R_*$. This value seems too large if one takes the Sobolev length as the typical clump size at the wind base, of $\sim (v_{\rm th}/V_{\rm w})R_*\sim 0.01\,R_*$, where $v_{\rm th}$ is the stellar wind ion thermal speed \cite[see e.g.][]{2006ApJ...648..565O}. The typical or {\it average} clump size at the two-wind interaction region then becomes $R_{\rm c,0}\sim 0.1\,d_1R_*$, hence to fulfill Eq.~(\ref{rccwr}) would require $f_{-1}^{1/2}\eta_{-1}^{1/2}\lesssim 0.1$. This would only be possible if $f$ were much smaller than the values $\sim 0.1$ typically considered, and/or $\eta$ were very small, which would imply a very weak pulsar wind and therefore a faint non-thermal source. Furthermore, clumps originating in the stellar wind acceleration region could be disrupted by Rayleigh-Taylor or thin-shell instabilities, limiting the clump growth when travelling away from the star \citep{2003A&A...406L...1D,2005A&A...437..657D}. It is worth noting that large clump-like structures may still form in the stellar wind for reasons different from instabilities occurring during the line-driven acceleration of the wind. In particular, the stellar rotation, magnetic field inhomogeneities in the stellar surface, or non-radial pulsations, may lead to clump-like, large-scale structures in the wind of size $\sim R_*$, like those inferred from the observed DACs in the UV. It is unclear whether the density contrast behind these observational UV features could be as large as $f^{-1}\sim 10$ \citep{1996ApJ...462..469C,2008ApJ...678..408L}. Nevertheless, the complexity of the stellar wind dynamics prevents us from discarding this possibility, as well as that of sporadic large and/or dense clumps of wind-acceleration region origin.

\subsection{Clump-pulsar wind interaction rates}\label{clsd}

An estimate of how often clumps fulfilling Eq.~(\ref{rccwr}) reach the two-wind interaction region can be calculated regardless of the clump origin. One can determine the duty cycle ($DC$) of these events by deriving their frequency ($\dot{N}$) and multiplying it by their duration $t_{\rm c}$, and also by adopting an empirical clump mass distribution without specifying the physical origin of those clumps.

To find $\dot{N}$, one needs first the number density distribution of clumps with radius
\begin{equation}
\frac{dN}{dR_{\rm c}dV}=n(R_{\rm c})\,.
\end{equation}
Assuming that all clumps are spherical and have the same density, the distribution function should meet the following normalization condition:
\begin{equation}
\frac{4\pi}{3}\int\limits_0^\infty R_{\rm c}^3 n(R_{\rm c})dR_{\rm c}=f\,.
\label{eq:dist_norm}
\end{equation}
Adopting for simplicity $n\propto R_{\rm c}^{-\alpha}$ for $R_{\rm c}>R_{\rm c,0}$, and $0$ otherwise, Eq.~(\ref{eq:dist_norm}) yields
\begin{equation}
n(R_c)=\frac{3f(\alpha-4)}{4\pi}R_{c,0}^{\alpha-4}R_c^{-\alpha}\,,  
\end{equation}
and the number density of clumps exceeding the limit given by Eq.~(\ref{rccwr}) is
\begin{equation}
\Upsilon=\int\limits_{f^{1/2}R}^\infty
n(R_{\rm c})dR_{\rm c}=\frac{3f(\alpha-4)}{4\pi(\alpha-1)}R_{\rm c,0}^{\alpha-4}\left(f^{1/2}R\right)^{1-\alpha}\,.
\end{equation}
We note that the derivation is valid for $\alpha>4$, since otherwise the integration in Eq.~(\ref{eq:dist_norm}) would diverge. This $\alpha$-range also means that the wind mass is mostly in the smallest clumps, of size $\sim R_{\rm c,0}$.

A large enough clump can strongly affect the emission from the two-wind interaction region if it propagates in the direction of the pulsar within a solid angle $\Delta\Omega=\pi (R_{\rm c}/R)^2\approx \pi f\eta$. Therefore, the frequency of arrival of large clumps to the vicinity of the pulsar can be estimated as
\begin{equation}
\dot{N}=\Delta \Omega d^2 V_{\rm w} \Upsilon\,,
\end{equation}
which allows the derivation of $DC$:
\begin{equation}
DC=t_{\rm c}\dot{N}=f^{-1/2}R_{\rm c} \Delta \Omega d^2 \Upsilon\,.
\end{equation}
We note that this is only approximately true because $t_{\rm c}$ is a function of clump size and $\dot{N}$ is an average. Taking an empirical mass power-law index $\alpha=5$, $t_{\rm c}=f^{1/2}R/V_{\rm w}$, and $V_{\rm w}=2\times 10^8$~cm~s$^{-1}$ (typical for massive star winds; \citealt{2012A&A...537A..37M}), one can obtain $DC=3/16\,(R_{\rm c,0}/R)\approx 0.006\eta_{-1}^{-1/2}$ and the typical time between large clump arrivals $\tau_{\rm c}=\dot{N}^{-1}\approx 2\times 10^6\eta_{-1}d_1$~s.

Clumps fulfilling Eq.~(\ref{cond2}) are expected to be rare, implying that their interaction with the pulsar wind should appear as a distinguishable event. Nevertheless, albeit unlikely, it cannot be discarded that clumps fulfilling Eqs.~(\ref{cond1}) or even (\ref{cond2}) could have $\tau_{\rm c}\lesssim t_{\rm c}$, in which case the simultaneous arrival of many clumps would smooth out the associated variability \citep[see also][]{2009ApJ...696..690O}.

\section{Radiation from clumpy stellar/pulsar wind collisions}\label{rad}

\subsection{Main characteristics of the clump-pulsar wind shock radiation}

In the clump-pulsar wind region two shocks form. One in the pulsar wind, and one in the clump. 
Non-thermal high-energy radiation, of a synchrotron and inverse Compton (IC) nature, is expected to originate in the shocked pulsar wind.
Relativistic bremsstrahlung is negligible in this light flow. The shock in the much denser clump has a luminosity smaller by 
$\sim f^{1/2}V_{\rm w}/2\,c$ 
than that of the pulsar wind shock. This is much smaller than unity, and implies that particle acceleration within the clump will be rather inefficient. 
The highest energy electrons may still penetrate in the clump from the shocked pulsar wind through diffusion and radiate there, but to characterize this process, fast radiative and non-radiative losses and diffusive transport should be carefully accounted for. Protons of pulsar wind origin have not been considered in this work.

\subsubsection{Dynamics and particle cooling}\label{dynrad}

Two extreme cases illustrate the impact of a clump interacting with the two-wind interaction region: (i) $X\gtrsim\Delta R$ and $R'\sim R$, or (ii) $R'\sim R_{\rm c}$. 

For case (i), $B$ will grow by a factor of $\sim (1+X/R)$ in the clump-pulsar wind interaction region. Given the relation between $B$ and the synchrotron cooling rate $\dot{E}_{\rm sync}\propto B^2$, the cooling rate will increase by $\sim (1+2X/R)$. Assuming an injection non-thermal luminosity $\sim (R_{\rm c}/R)^2(\chi\,L_{\rm sd})$ in the clump-pulsar wind interaction region, where $\chi\,L_{\rm sd}$ is the total non-thermal luminosity, the maximum variation of the synchrotron emission will be $\sim [1+(R_{\rm c}/R)^2(2X/R)]$. For $R_{\rm c}=0.3\,R_*$, $f=0.1$, $\eta=0.1$, and $d=10\,R_*$ (so $R\sim 3\,R_*$), the expected fluctuations will be at the level of $\sim 10$\%. The variations in IC and non-radiative cooling will be similar or even smaller. 

As $R'$ becomes $\lesssim 2\,R_{\rm c}$ (case ii), the situation changes quickly. From that point on, a substantial fraction of the pulsar wind luminosity crosses the clump-pulsar wind shock, and synchrotron cooling is now enhanced by a factor of $\sim (R/R')^2$ (as $B\propto R/R'$). On the other hand, IC cooling only changes slightly for $\eta\ll 1$. Non-radiative cooling grows by $\sim (R/R')$. As the loss balance changes, radiation is re-distributed in photon energy.

\subsubsection{Non-thermal emission}\label{nt}

The rise time of the lightcurve of the emission produced during a clump-pulsar wind interaction is linked to the non-radiative timescale (see Sect.~\ref{phfr}):
\begin{equation}
t_{\rm ad}\approx 170\,(R_{\rm c}/R_*)\,{\rm s};
\end{equation}
to the acceleration timescale ($B_{\rm G}=B/1\,{\rm G}$, $E_{\rm TeV}=E/1\,{\rm TeV}$; with $E$ the electron energy and $\xi_{0.1}=\xi/0.1$ the acceleration efficiency):
\begin{equation}
t_{\rm acc}\approx 1\,E_{\rm TeV}\xi_{0.1}^{-1}B_{\rm G}^{-1}\,{\rm s};
\end{equation}
to the synchrotron timescale: 
\begin{equation}
t_{\rm sync}\approx 400\,B_{\rm G}^{-1}\,E_{\rm TeV}^{-1}\,{\rm s}\,;
\end{equation}
to the IC timescale: 
\begin{equation}
t_{\rm IC}\approx 160\,u_2^{-1}\,E_{\rm GeV}^{-1}\,{\rm s}
\end{equation}
in the Thomson regime ($u_{2}=u/100\,{\rm erg~cm}^{-3}$, $E_{\rm GeV}=E/1\,{\rm GeV}$), and
\begin{equation}
t_{\rm IC}\approx 100\,E_{\rm TeV}^{0.7}\,T_{4.5}\,u_2^{-1}\,{\rm s},
\end{equation}
in the Klein-Nishina regime ($T_{4.5}=T/3\times 10^4{\rm K}$) \citep[for radiation and acceleration timescales, see][]{2008MNRAS.383..467K}; 
and to the clump expansion timescale:
\begin{equation}
\sim t_{\rm c}\sim 10^4f_{-1}^{-1/2}(R_{\rm c}/R_*)\,{\rm s}.
\end{equation}
Since the particle energy gain and loss timescales will be generally much shorter than $t_{\rm c}$, the accelerated particles will reach the steady state, or an adiabatic succession of quasi-steady states, before the clump gets destroyed by the pulsar wind. The lightcurve will therefore be strongly linked to the clump dynamics.
 
Clump expansion is rapid and occurs simultaneously with clump braking and deflection by the pulsar wind. As illustrated in Fig.~\ref{sketch}, the clump ends up as a tail of shocked material that will join the tail of the whole two-wind interaction region. For an accurate modelling of the lightcurve, the complexity of the clump destruction deserves a detailed treatment that is beyond the scope of this work. Nevertheless, we note that for the case $R'>R_{\rm c}\gtrsim \Delta R$, the emission from the clump-pulsar wind interaction can be additionally enhanced by a factor of a few during the clump expansion with a sharpening of the lightcurve \citep[see e.g.][and Sect.~\ref{disc}, for the cloud radius evolution]{2010ApJ...724.1517B,2012A&A...539A..69B}. 

The balance between the different cooling channels changes with the clump arrival, which, putting aside non-radiative cooling, diverts radiation energy from the highest energies (IC) to lower ones (synchrotron). The synchrotron and the IC luminosity can be approximately described by
\begin{equation}
L_{\rm sync,IC,0}\approx (L_{\rm sd}/2)\frac{t_{\rm sync,IC,0}^{-1}}{t_{\rm sync,0}^{-1}+t_{\rm IC}^{-1}+t_{\rm ad,0}^{-1}}
\end{equation}
before the clump arrival. After the arrival of a clump with $R_{\rm c}\sim f^{1/2}R$, the clump-pulsar wind interaction region emits synchrotron radiation as
\begin{equation}
L_{\rm sync}\approx (L_{\rm sd}/2)
\times\frac{(R/R')^2t_{\rm sync,0}^{-1}}{(R/R')^2t_{\rm sync,0}^{-1}+t_{\rm IC,0}^{-1}+(R/R')t_{\rm ad,0}^{-1}}
\end{equation}
and IC radiation as
\begin{equation}
L_{\rm IC}\approx (L_{\rm sd}/2)
\times\frac{t_{\rm IC,0}^{-1}}{(R/R')^2t_{\rm sync,0}^{-1}+t_{\rm IC,0}^{-1}+(R/R')t_{\rm ad,0}^{-1}}\,.
\end{equation}
The cooling balance shows that, when a large clump reaches the pulsar vicinity, the synchrotron emission grows in the clump-pulsar wind interaction region unless it was the dominant cooling channel before the clump arrival. On the other hand, the IC emission always diminishes. Non-radiative losses, which grow as well but more slowly than synchrotron radiation, moderate when they are dominant the $R'$-dependence of the two radiation channels.

Inverse Compton scattering of electrons with their own synchrotron photons may also become important as the emitter size decreases from $R$ to $R'\sim R_{\rm c}$. The same applies to gamma-ray absorption through pair creation in the synchrotron field. One can estimate the influence of size reduction comparing IC scattering with stellar photons and synchrotron self-Compton (SSC) in the Thomson regime approximation, the former being dominant unless $f\eta\lesssim L_{\rm sd}/L_*$, in which case the internal photon energy density can overcome that of the star. This requirement can be rewritten as 
\begin{equation}
L_{\rm sd}>10^{36}\eta_{-1}f_{-1}L_{*38}\,{\rm erg~s}^{-1},
\end{equation}
or
\begin{equation}
\dot{M}V_{\rm w}c>10^{37}f_{-1}L_{*38}\,{\rm erg~s}^{-1},
\end{equation}
where $\dot{M}$ is the wind mass-loss rate and $L_{*38}=(L_*/10^{38}\,{\rm erg~s}^{-1})$, and may be met in sources such as PSR~B1259$-$63. A similar condition applies for significant pair creation in the synchrotron field given that the pair-creation cross section is similar to that of IC scattering in the Klein-Nishina regime. The stellar and the synchrotron spectra are however different, and thus the transition between the Thomson and the Klein-Nishina regime \citep[e.g.][]{1970RvMP...42..237B} occurs differently in both target fields. This introduces some uncertainty in the estimated constraint, but given that the synchrotron photons in the region are more energetic than the stellar photons, the constraint would actually be stricter.

\subsection{Semi-analytic modelling of an illustrative case}\label{semian}

\begin{figure}
\vspace{0.7cm}
\includegraphics[width=0.44\textwidth,angle=0]{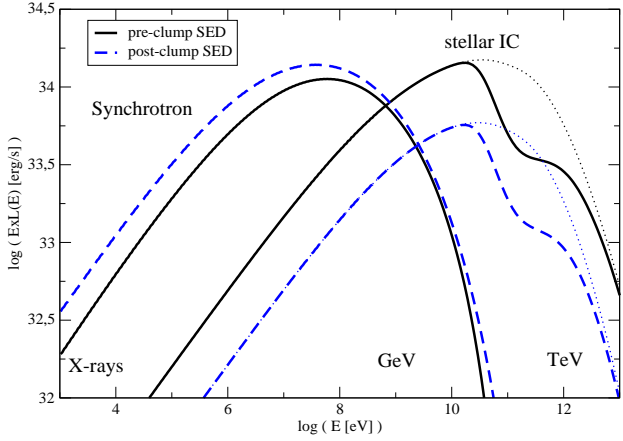}
\vspace{0.3cm}
\caption{Spectral energy distribution of the synchrotron and the absorbed (thick lines) and unabsorbed (thin dotted lines) stellar IC emission before and after the clump arrival for the case with $\sigma=0.01$.}
\label{lowb}
\end{figure}

In this section, we present the results of semi-analytical calculations of the synchrotron, synchrotron self-Compton, stellar photon IC emission, and external (binary) and internal (emitter) pair creation and its subsequent pair radiation. For these calculations, we focused on the case with $d_{1}=1$, $\eta_{-1}=1$, $f_{-1}=1$, $R_{\rm c}=R_*$, all this yielding $R'\sim R_{\rm c}\approx R/3$. We also took $R_*=10\,R_\odot$, and fixed $L_{\rm sd}$ to $10^{36}$~erg~s$^{-1}$, taking it as equal to the injected non-thermal power (i.e. $\chi=1$). We also adopted two magnetic-to-total energy density ratios in the shocked pulsar wind of $\sigma=0.01$ and 1, giving an initial magnetic field in the two-wind interaction region of $B\approx 0.3$ and 3~G. Therefore, after the clump arrival, $B$ becomes $\approx 1$ and 10~G in the clump-pulsar wind interaction region; these two magnetic field values were adopted as fiducial values for a kinetic and a magnetic pulsar wind, respectively. We took an acceleration efficiency in the clump-pulsar wind shock of $\xi=0.1$ to obtain the maximum particle energy $E_{\rm max}$, with an injection energy distribution $\propto E^{-2}\exp{(-E/E_{\rm max})}$. The emitter was assumed to be homogeneous (one zone), with size $R$ and $R'$ and a non-radiative timescale $\sqrt{3}R/c$ and $\sqrt{3}R'/c$ before and after the clump arrival. The stellar luminosity and temperature were taken $3\times 10^{38}$~erg~s$^{-1}$ and $3\times 10^4$~K, respectively, with the star being considered a black body and point-like.

To avoid the complexities of IC scattering and gamma-ray absorption from the orbital motion and the emitter-observer geometry relations \citep[e.g.][]{2008MNRAS.383..467K}, not relevant at this stage, we adopted an isotropic stellar field. Moreover, we computed the secondary emission in the one-zone approximation. For the pairs created within the system, we took an emitter of size $d$ with a radiation energy density fixed to its value at a distance $d$ from the star. The $B$-value in the system was taken 10 times below equipartition with the stellar photon field. However, as expected (see Sect.~\ref{nt}), both internal and external absorption yielded a relatively small number of pairs. Therefore, secondary emission, as well as SSC, well below the external IC component, are not considered further. 

\begin{figure}
\vspace{0.7cm}
\includegraphics[width=0.44\textwidth,angle=0]{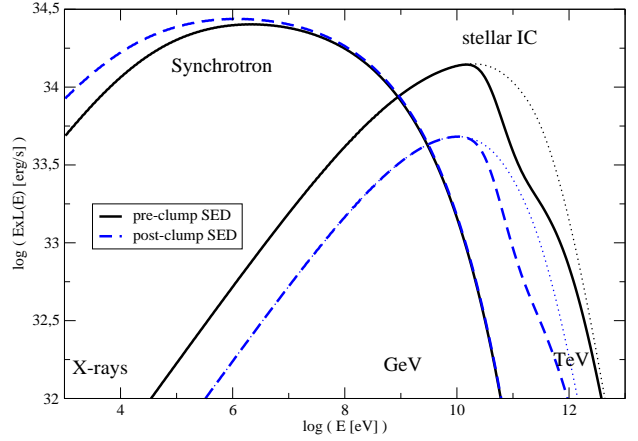}
\vspace{0.3cm}
\caption{The same as in Fig.~\ref{lowb} but for $\sigma=1$.}
\label{highb}
\end{figure}

We present in Figs.~\ref{lowb} and \ref{highb} the broadband spectral energy distribution (SED) of the synchrotron and stellar IC radiation with gamma-ray absorption by the stellar photons. 
The pre- and post-clump SEDs are presented together, each figure corresponding to a different magnetic field: $\sigma=0.01$ (Fig.~\ref{lowb}) and 1 (Fig.~\ref{highb}). Results are similar for both cases, although the radiative cooling regime starts at higher energies for the lower $\sigma$ case. 
Below the highest energies of the synchrotron and IC components and for the adopted parameters, non-radiative cooling shapes the SED, so the whole synchrotron luminosity grows $\propto (R/R')\approx 3$ and the IC emission decreases by the same factor after the clump arrival. At high photon energies of both radiation components, synchrotron cooling shapes the SED before and after the clump arrival in both cases, and therefore the synchrotron luminosity does not change, whereas the IC radiation decreases $\propto (R'/R)^2$ (hard to see because of the steepness of the SED). 
We note that for a fixed energy band, synchrotron emission is actually $\propto B^{1.5}$ (for a spectral index $\alpha=0.5$, where $F_\nu\propto \nu^{-\alpha}$), so when balanced by non-radiative cooling 
one gets an X-ray flux enhancement $\propto B^{0.5}$, i.e. a factor of $\approx 2$ for $f=0.1$. This does not occur in IC emission.
Another spectral feature is that the SED tends to soften for larger magnetic fields.

For the studied case, non-radiative losses reduce the synchrotron enhancement in X-rays from $\sim 10$ to a factor of a few. The reduction of IC emission at GeV energies is also moderated by the dominant non-radiative losses. Given that non-radiative losses are $\propto d^{-1}$, and synchrotron and IC losses $\propto d^{-2}$, for more compact binaries radiation cooling gets more prominent, and the contrast between luminosities before and after the clump arrival becomes higher at lower energies. For wide systems, on the other hand, non-radiative losses can be dominant in the whole particle energy range. 

\section{Discussion and summary}\label{disc}

The structure of stellar winds in massive binaries hosting a powerful pulsar can have significant effects in the non-thermal lightcurve and spectrum. The two dominant factors are the increase in $B$, which directly affects the synchrotron emission and indirectly, the IC emission, and the shorter non-radiative cooling time. Nevertheless, the synchrotron losses grow faster, $\propto R'^{-2}$, than the non-radiative losses, $\propto R'^{-1}$. Inverse Compton cooling is only slightly affected for $\eta\ll 1$. 

An open issue is whether clumps large and/or dense enough to penetrate deeply in the pulsar wind region can form in the stellar wind. Instabilities in the stellar wind acceleration region could be a possibility, but the related clumps may be too small if their growth is limited by clump disruption beyond some size. The large-scale stellar wind structures behind DACs may be another possibility, although it is unclear whether the density contrast in this case is enough to fulfill the pulsar wind deep-penetration condition.

Regardless of the physical origin of the clumps, if the number density distribution of clumps with radius can be approximated by an empirical power-law at least up to $R_{\rm c}\sim f_{-1}^{1/2}\eta_{-1}^{1/2}d_{1}R_*$, some clumps can sporadically collide with the two-wind interaction region inducing variations in the emission by a factor of several on timescales of minutes to hours (more on clump expansion below). 
The duty cycles strongly depend on the structure of the stellar wind, but duty cycles of $\sim 1$\% seem feasible.

There are other issues that deserve discussion in the studied scenario. We comment on a number of them in what follows: (i) the clump expansion; (ii) the Doppler boosting of the clump radiation when the shocked clump material approaches relativistic speeds; (iii) the effect of clump fragmentation; (iv) the magnetic field; (v) the clump thermal radiation; and (vi) observations. 

\begin{figure}
\vspace{0.6cm}
\includegraphics[width=0.42\textwidth,angle=0]{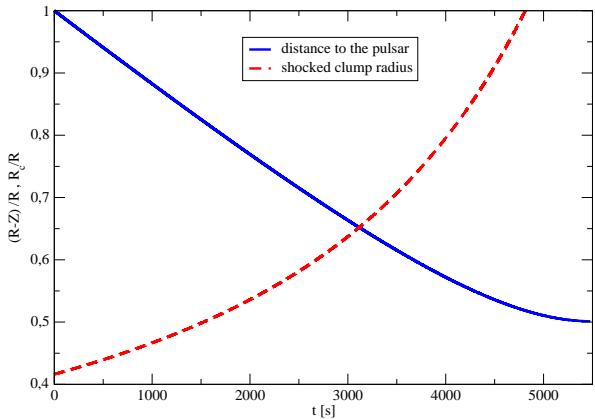}
\vspace{0.3cm}
\caption{Distance to the pulsar (solid line) and shocked clump radius (dashed line) normalized to the size of the two-wind interaction region.}
\label{ex}
\end{figure}

Clump expansion (i) in the vicinity of the pulsar can be very fast. Using Eqs.~(9)-(13) in \cite{2010ApJ...724.1517B} and accounting for the radial geometry of the pulsar wind, we calculated the cloud expansion until the cloud advance is stopped by the pulsar wind, which result is shown in Fig.~\ref{ex}. As seen in the figure, $R_{\rm c}$ reaches the size of the whole two-wind interaction region around the point where the clump stops. The result is consistent with the shocked clump material eventually forming its own bow-shaped structure. The calculation also shows that the clump expansion accelerates, and its last stages may take minutes after an initial phase lasting about one hour, under the adopted parameter values. These last stages of the expansion may correspond to a sharpening of the lightcurve.
The value of $R'$ found in the calculation is slightly different to that derived analytically adopting the same parameters. However, the calculation of the cloud evolution has been largely simplified, as instabilities and clump fragmentation have been neglected. More detailed numerical simulations of a relativistic outflow impacting a spherical cloud \citep[e.g.][]{2012A&A...539A..69B} show that the clump expansion rate, albeit high, is somewhat slower than the one computed here, and more in line with the analytical estimates. 

Regarding Doppler boosting (ii), the clump expansion and propagation velocities do not become relativistic before dissipation, although the shocked pulsar wind circumventing the clump should have an accelerating relativistic speed \citep{2008MNRAS.387...63B}. Finer calculations should account for this effect, which can introduce line-of-sight dependencies, i.e. an additional variability factor.

The last stages of the clump-pulsar wind interaction may be accompanied by features in the lightcurve and the spectrum linked to cloud fragmentation (iii). Once disrupted, the clump material becomes a slow and dense inhomogeneous flow that joins the shocked two-wind structure and is advected by it. This will excite shocks, turbulence and two-wind mixing in the flow (as suggested, e.g., in \citealt{2010MNRAS.403.1873Z,2012A&A...546A..60L}), which are in fact already present when a smooth stellar wind interacts with the relativistic pulsar wind \citep{2012A&A...544A..59B}.

The clump propagation through the shocked stellar and pulsar wind can strongly interfere in the magnetic field advected downstream (iv). The $B$-lines will be warped and stretched by the clump, which will lead to $B$-growth, instabilities, turbulence, and reconnection in the plasma trailing the clump \citep[e.g.][]{1996ApJ...473..365J}; all these effects are potentially relevant for the non-thermal activity around the clump. 

Modest thermal emission (v) in the form of soft X-ray lines and thermal Bremsstrahlung is expected from the clump because of the impact of the pulsar wind. Initially, the shocked clump produces thermal photons $f$ times cooler than those of the shocked smooth wind, which mainly emits in the soft X-ray band \citep[see e.g.][]{2011ApJ...743....7Z}. As clump heating and expansion proceed, the thermal photon energy reaches that of the shocked smooth wind. The clump gets then hotter but also significantly diluted, quenching its thermal radiation.

Certain observational features (vi) detected in binaries potentially hosting a pulsar may be explained by invoking the presence of clumps, for instance short flares of scales of seconds to hours found in the X-ray lightcurves of LS~5039 and LS~I~+61~303 \citep[e.g.][]{2005ApJ...628..388B,2007ApJ...664L..39P,2009ApJ...693.1621S,2011ApJ...733...89L}. The observational hints of stellar wind inhomogeneities, manifesting themselves mostly through short X-ray variability, and the complex evolution of the wind structures under the impact of the pulsar wind, call for coupled dynamics-radiation studies and their comparison with observations. Besides improving our understanding of the non-thermal processes within the binary, these studies can also probe the stellar wind properties in high-mass binaries hosting young pulsars.

We want to emphasize that direct attempts to simulate the processes considered approximately in this work are very necessary, since as noted an analytic treatment does not allow for precise quantitative predictions and may miss subtle effects of major impact. Despite our conclusions pointing to a smooth effect of moderate clumping on the two-wind post-shock regions \citep[see also][]{2007ApJ...660L.141P}, the relativistic nature of the pulsar wind \citep[about the related complexities, see e.g. the discussion in][]{2012A&A...544A..59B}, and the range of possible interaction parameters of the sources of interest, open the door to unexpected effects even when only modest size clumps could be present in the stellar wind. In addition, the presence of additional larger inhomogeneities in the stellar wind (like those behind DACs) could provide a strong source of variability, and also introduce substantial changes not only in the size, but also in the orientation and the overall structure of the two-wind interaction region. Finally, it is noteworthy that PSR~B1259$-$63, the only confirmed high-mass binary hosting a powerful pulsar so far, has a circumstellar disc, which adds further complications to the evolution of the two-wind interaction structure \citep[see e.g.][]{2011PASJ...63..893O,2012ApJ...750...70T}. In short, numerical three-dimensional relativistic hydrodynamical simulations of a stellar wind interacting with a pulsar wind, accounting for both modest and large wind structures, are needed for a more accurate assessment of the required clump size to have a significant observational impact. This should be done at least on a spatial scale $\sim d$, and the inclusion of a magnetic field would be useful. The required size of the grid plus a minimum resolution of a few cells per $R_{\rm c}$ mean a rather heavy calculation, although box-type simulations on spatial scales $\sim R$ could already be quite informative.
Detailed clump time-arrival calculations, as done in \cite{2012MNRAS.421.2820O}, would be also informative to interpret lightcurves.

\begin{acknowledgements} 
We want to thank Stan Owocki for a constructive and really useful review.
We also thank Dmitry Khangulyan for his useful comments on the manuscript.
We acknowledge support by DGI of the Spanish Ministerio de Econom\'{\i}a
y Competitividad (MINECO) under grants  AYA2010-21782-C03-01 and FPA2010-22056-C06-02.
V.B-R. acknowledges financial support from MINECO through a Ram\'on y
Cajal fellowship. This research has been
supported by the Marie Curie Career Integration Grant 321520.
\end{acknowledgements}

\bibliographystyle{aa}
\bibliography{text}

\begin{thebibliography}{49}
\expandafter\ifx\csname natexlab\endcsname\relax\def\natexlab#1{#1}\fi

\bibitem[{{Araudo} {et~al.}(2009){Araudo}, {Bosch-Ramon}, \&
  {Romero}}]{2009A&A...503..673A}
{Araudo}, A.~T., {Bosch-Ramon}, V., \& {Romero}, G.~E. 2009, \aap, 503, 673

\bibitem[{{Barkov} {et~al.}(2010){Barkov}, {Aharonian}, \&
  {Bosch-Ramon}}]{2010ApJ...724.1517B}
{Barkov}, M.~V., {Aharonian}, F.~A., \& {Bosch-Ramon}, V. 2010, \apj, 724, 1517

\bibitem[{{Blumenthal} \& {Gould}(1970)}]{1970RvMP...42..237B}
{Blumenthal}, G.~R. \& {Gould}, R.~J. 1970, Reviews of Modern Physics, 42, 237

\bibitem[{{Bogovalov} {et~al.}(2008){Bogovalov}, {Khangulyan}, {Koldoba},
  {Ustyugova}, \& {Aharonian}}]{2008MNRAS.387...63B}
{Bogovalov}, S.~V., {Khangulyan}, D.~V., {Koldoba}, A.~V., {Ustyugova}, G.~V.,
  \& {Aharonian}, F.~A. 2008, \mnras, 387, 63

\bibitem[{{Bosch-Ramon}(2010)}]{2010tsra.confE.179B}
{Bosch-Ramon}, V. 2010, in 25th Texas Symposium on Relativistic Astrophysics

\bibitem[{{Bosch-Ramon} {et~al.}(2012{\natexlab{a}}){Bosch-Ramon}, {Barkov},
  {Khangulyan}, \& {Perucho}}]{2012A&A...544A..59B}
{Bosch-Ramon}, V., {Barkov}, M.~V., {Khangulyan}, D., \& {Perucho}, M.
  2012{\natexlab{a}}, \aap, 544, A59

\bibitem[{{Bosch-Ramon} {et~al.}(2005){Bosch-Ramon}, {Paredes}, {Rib{\'o}},
  {Miller}, {Reig}, \& {Mart{\'{\i}}}}]{2005ApJ...628..388B}
{Bosch-Ramon}, V., {Paredes}, J.~M., {Rib{\'o}}, M., {et~al.} 2005, \apj, 628,
  388

\bibitem[{{Bosch-Ramon} {et~al.}(2012{\natexlab{b}}){Bosch-Ramon}, {Perucho},
  \& {Barkov}}]{2012A&A...539A..69B}
{Bosch-Ramon}, V., {Perucho}, M., \& {Barkov}, M.~V. 2012{\natexlab{b}}, \aap,
  539, A69

\bibitem[{{Cranmer} \& {Owocki}(1996)}]{1996ApJ...462..469C}
{Cranmer}, S.~R. \& {Owocki}, S.~P. 1996, \apj, 462, 469

\bibitem[{{Dessart} \& {Owocki}(2003)}]{2003A&A...406L...1D}
{Dessart}, L. \& {Owocki}, S.~P. 2003, \aap, 406, L1

\bibitem[{{Dessart} \& {Owocki}(2005)}]{2005A&A...437..657D}
{Dessart}, L. \& {Owocki}, S.~P. 2005, \aap, 437, 657

\bibitem[{{Dubus}(2006)}]{2006A&A...456..801D}
{Dubus}, G. 2006, \aap, 456, 801

\bibitem[{{Dubus}(2013)}]{2013arXiv1307.7083D}
{Dubus}, G. 2013, ArXiv e-prints

\bibitem[{{Hirschi}(2008)}]{2008cihw.conf....9H}
{Hirschi}, R. 2008, in Clumping in Hot-Star Winds, ed. W.-R. {Hamann},
  A.~{Feldmeier}, \& L.~M. {Oskinova}, 9

\bibitem[{{Johnston} {et~al.}(1992){Johnston}, {Manchester}, {Lyne}, {Bailes},
  {Kaspi}, {Qiao}, \& {D'Amico}}]{1992ApJ...387L..37J}
{Johnston}, S., {Manchester}, R.~N., {Lyne}, A.~G., {et~al.} 1992, \apjl, 387,
  L37

\bibitem[{{Jones} {et~al.}(1996){Jones}, {Ryu}, \&
  {Tregillis}}]{1996ApJ...473..365J}
{Jones}, T.~W., {Ryu}, D., \& {Tregillis}, I.~L. 1996, \apj, 473, 365

\bibitem[{{Khangulyan} {et~al.}(2008){Khangulyan}, {Aharonian}, \&
  {Bosch-Ramon}}]{2008MNRAS.383..467K}
{Khangulyan}, D., {Aharonian}, F., \& {Bosch-Ramon}, V. 2008, \mnras, 383, 467

\bibitem[{{Khangulyan} {et~al.}(2007){Khangulyan}, {Hnatic}, {Aharonian}, \&
  {Bogovalov}}]{2007MNRAS.380..320K}
{Khangulyan}, D., {Hnatic}, S., {Aharonian}, F., \& {Bogovalov}, S. 2007,
  \mnras, 380, 320

\bibitem[{{Kirk} {et~al.}(1999){Kirk}, {Ball}, \&
  {Skjaeraasen}}]{1999APh....10...31K}
{Kirk}, J.~G., {Ball}, L., \& {Skjaeraasen}, O. 1999, Astroparticle Physics,
  10, 31

\bibitem[{{Lamberts} {et~al.}(2012){Lamberts}, {Dubus}, {Lesur}, \&
  {Fromang}}]{2012A&A...546A..60L}
{Lamberts}, A., {Dubus}, G., {Lesur}, G., \& {Fromang}, S. 2012, \aap, 546, A60

\bibitem[{{Lamberts} {et~al.}(2013){Lamberts}, {Fromang}, {Dubus}, \&
  {Teyssier}}]{2013arXiv1309.7629L}
{Lamberts}, A., {Fromang}, S., {Dubus}, G., \& {Teyssier}, R. 2013, ArXiv
  e-prints

\bibitem[{{Li} {et~al.}(2011){Li}, {Torres}, {Zhang}, {Chen}, {Hadasch}, {Ray},
  {Kretschmar}, {Rea}, \& {Wang}}]{2011ApJ...733...89L}
{Li}, J., {Torres}, D.~F., {Zhang}, S., {et~al.} 2011, \apj, 733, 89

\bibitem[{{Lobel} \& {Blomme}(2008)}]{2008ApJ...678..408L}
{Lobel}, A. \& {Blomme}, R. 2008, \apj, 678, 408

\bibitem[{{Lucy} \& {Solomon}(1970)}]{1970ApJ...159..879L}
{Lucy}, L.~B. \& {Solomon}, P.~M. 1970, \apj, 159, 879

\bibitem[{{Maraschi} \& {Treves}(1981)}]{1981MNRAS.194P...1M}
{Maraschi}, L. \& {Treves}, A. 1981, \mnras, 194, 1P

\bibitem[{{Moffat}(2008)}]{2008cihw.conf...17M}
{Moffat}, A.~F.~J. 2008, in Clumping in Hot-Star Winds, ed. W.-R. {Hamann},
  A.~{Feldmeier}, \& L.~M. {Oskinova}, 17

\bibitem[{{Mold{\'o}n} {et~al.}(2011){Mold{\'o}n}, {Johnston}, {Rib{\'o}},
  {Paredes}, \& {Deller}}]{2011ApJ...732L..10M}
{Mold{\'o}n}, J., {Johnston}, S., {Rib{\'o}}, M., {Paredes}, J.~M., \&
  {Deller}, A.~T. 2011, \apjl, 732, L10

\bibitem[{{Muijres} {et~al.}(2012){Muijres}, {Vink}, {de Koter}, {M{\"u}ller},
  \& {Langer}}]{2012A&A...537A..37M}
{Muijres}, L.~E., {Vink}, J.~S., {de Koter}, A., {M{\"u}ller}, P.~E., \&
  {Langer}, N. 2012, \aap, 537, A37

\bibitem[{{Negueruela} {et~al.}(2011){Negueruela}, {Rib{\'o}}, {Herrero},
  {Lorenzo}, {Khangulyan}, \& {Aharonian}}]{2011ApJ...732L..11N}
{Negueruela}, I., {Rib{\'o}}, M., {Herrero}, A., {et~al.} 2011, \apjl, 732, L11

\bibitem[{{Neronov} \& {Chernyakova}(2007)}]{2007Ap&SS.309..253N}
{Neronov}, A. \& {Chernyakova}, M. 2007, \apss, 309, 253

\bibitem[{{Okazaki} {et~al.}(2011){Okazaki}, {Nagataki}, {Naito}, {Kawachi},
  {Hayasaki}, {Owocki}, \& {Takata}}]{2011PASJ...63..893O}
{Okazaki}, A.~T., {Nagataki}, S., {Naito}, T., {et~al.} 2011, \pasj, 63, 893

\bibitem[{{Oskinova} {et~al.}(2012){Oskinova}, {Feldmeier}, \&
  {Kretschmar}}]{2012MNRAS.421.2820O}
{Oskinova}, L.~M., {Feldmeier}, A., \& {Kretschmar}, P. 2012, \mnras, 421, 2820

\bibitem[{{Owocki} \& {Cohen}(2006)}]{2006ApJ...648..565O}
{Owocki}, S.~P. \& {Cohen}, D.~H. 2006, \apj, 648, 565

\bibitem[{{Owocki} {et~al.}(2009){Owocki}, {Romero}, {Townsend}, \&
  {Araudo}}]{2009ApJ...696..690O}
{Owocki}, S.~P., {Romero}, G.~E., {Townsend}, R.~H.~D., \& {Araudo}, A.~T.
  2009, \apj, 696, 690

\bibitem[{{Papitto} {et~al.}(2012){Papitto}, {Torres}, \&
  {Rea}}]{2012ApJ...756..188P}
{Papitto}, A., {Torres}, D.~F., \& {Rea}, N. 2012, \apj, 756, 188

\bibitem[{{Paredes} {et~al.}(2013){Paredes}, {Bednarek}, {Bordas},
  {Bosch-Ramon}, {De Cea del Pozo}, {Dubus}, {Funk}, {Hadasch}, {Khangulyan},
  {Markoff}, {Mold{\'o}n}, {Munar-Adrover}, {Nagataki}, {Naito}, {de Naurois},
  {Pedaletti}, {Reimer}, {Rib{\'o}}, {Szostek}, {Terada}, {Torres}, {Zabalza},
  {Zdziarski}, \& {CTA Consortium}}]{2013APh....43..301P}
{Paredes}, J.~M., {Bednarek}, W., {Bordas}, P., {et~al.} 2013, Astroparticle
  Physics, 43, 301

\bibitem[{{Paredes} {et~al.}(2007){Paredes}, {Rib{\'o}}, {Bosch-Ramon}, {West},
  {Butt}, {Torres}, \& {Mart{\'{\i}}}}]{2007ApJ...664L..39P}
{Paredes}, J.~M., {Rib{\'o}}, M., {Bosch-Ramon}, V., {et~al.} 2007, \apjl, 664,
  L39

\bibitem[{{Perucho} \& {Bosch-Ramon}(2012)}]{2012A&A...539A..57P}
{Perucho}, M. \& {Bosch-Ramon}, V. 2012, \aap, 539, A57

\bibitem[{{Pittard}(2007)}]{2007ApJ...660L.141P}
{Pittard}, J.~M. 2007, \apjl, 660, L141

\bibitem[{{Portegies Zwart} \& {Yungelson}(1999)}]{1999MNRAS.309...26P}
{Portegies Zwart}, S.~F. \& {Yungelson}, L.~R. 1999, \mnras, 309, 26

\bibitem[{{Puls} {et~al.}(2006){Puls}, {Markova}, {Scuderi}, {Stanghellini},
  {Taranova}, {Burnley}, \& {Howarth}}]{2006A&A...454..625P}
{Puls}, J., {Markova}, N., {Scuderi}, S., {et~al.} 2006, \aap, 454, 625

\bibitem[{{Runacres} \& {Owocki}(2002)}]{2002A&A...381.1015R}
{Runacres}, M.~C. \& {Owocki}, S.~P. 2002, \aap, 381, 1015

\bibitem[{{Smith} {et~al.}(2009){Smith}, {Kaaret}, {Holder}, {Falcone},
  {Maier}, {Pandel}, \& {Stroh}}]{2009ApJ...693.1621S}
{Smith}, A., {Kaaret}, P., {Holder}, J., {et~al.} 2009, \apj, 693, 1621

\bibitem[{{Takata} {et~al.}(2012){Takata}, {Okazaki}, {Nagataki}, {Naito},
  {Kawachi}, {Lee}, {Mori}, {Hayasaki}, {Yamaguchi}, \&
  {Owocki}}]{2012ApJ...750...70T}
{Takata}, J., {Okazaki}, A.~T., {Nagataki}, S., {et~al.} 2012, \apj, 750, 70

\bibitem[{{Tavani} \& {Arons}(1997)}]{1997ApJ...477..439T}
{Tavani}, M. \& {Arons}, J. 1997, \apj, 477, 439

\bibitem[{{Walter} \& {Zurita Heras}(2007)}]{2007A&A...476..335W}
{Walter}, R. \& {Zurita Heras}, J. 2007, \aap, 476, 335

\bibitem[{{Zabalza} {et~al.}(2013){Zabalza}, {Bosch-Ramon}, {Aharonian}, \&
  {Khangulyan}}]{2013A&A...551A..17Z}
{Zabalza}, V., {Bosch-Ramon}, V., {Aharonian}, F., \& {Khangulyan}, D. 2013,
  \aap, 551, A17

\bibitem[{{Zabalza} {et~al.}(2011){Zabalza}, {Bosch-Ramon}, \&
  {Paredes}}]{2011ApJ...743....7Z}
{Zabalza}, V., {Bosch-Ramon}, V., \& {Paredes}, J.~M. 2011, \apj, 743, 7

\bibitem[{{Zdziarski} {et~al.}(2010){Zdziarski}, {Neronov}, \&
  {Chernyakova}}]{2010MNRAS.403.1873Z}
{Zdziarski}, A.~A., {Neronov}, A., \& {Chernyakova}, M. 2010, \mnras, 403, 1873

\end{thebibliography}
\end{document}